\documentstyle[aps,preprint,epsfig,12pt]{revtex}

\begin{document}

\tighten

\preprint{
\noindent
\hfill
\begin{minipage}[t]{3in}
\begin{flushright}
TU-611\\
hep-ph/0102204\\
February 2001\\
\end{flushright}
\end{minipage}
}

\title{
\vskip 0.5in
Recent Result from E821 Experiment on Muon $g-2$ and 
Unconstrained Minimal Supersymemtric Standard Model
}

\author{Shinji Komine,\footnote{e-mail: komine@tuhep.phys.tohoku.ac.jp}
 Takeo Moroi\footnote{e-mail: moroi@tuhep.phys.tohoku.ac.jp} and 
Masahiro Yamaguchi\footnote{e-mail: yamaguchi@phys.tohoku.ac.jp}}

\address{Department of Physics, Tohoku University,
Sendai 980-8578, Japan}

\maketitle

\begin{abstract}

    Recently, the E821 experiment at the Brookhaven National
    Laboratory announced their latest result of their muon $g-2$
    measurement which is about 2.6-$\sigma$ away from the standard
    model prediction.  Taking this result seriously, we examine the
    possibility to explain this discrepancy by the supersymmetric
    contribution.  Our analysis is performed in the framework of the
    {\it unconstrained} supersymmetric standard model which has free
    seven parameters relevant to muon $g-2$.  We found that, in the
    case of large $\tan\beta$, sparticle masses are allowed to be
    large in the region where the SUSY contribution to the muon $g-2$
    is large enough, and hence the conventional SUSY search may fail
    even at the LHC.  On the contrary, to explain the discrepancy in
    the case of small $\tan\beta$, we found that (i) sleptons and
    $SU(2)_L$ gauginos should be light, and (ii) negative search for
    the Higgs boson severely constrains the model in the framework of
    the mSUGRA and gauge-mediated model.

\end{abstract} 

\clearpage

The latest result of the E821 experiment at Brookhaven National Laboratory 
on the muon
anomalous magnetic moment $g_{\mu}-2$ \cite{E821latest}
\begin{equation}
    a_{\mu}(\mbox{E821}) \equiv (g_{\mu}-2)/2 = 
    11 659 202 (14)(6) \times 10^{-10}
    \label{eq:MDMexp}
\end{equation} 
indicates a possible confrontation with the standard model of particle
physics. The deviation between the experimental value and the standard
model prediction is
\begin{equation}
    a_{\mu}(\mbox{E821}) - a_{\mu}(\mbox{SM}) =  43 (16) \times 10^{-10},
\label{eq:MDMdeviation}
\end{equation}
or 2.6-$\sigma$ deviation \cite{E821latest}. In order to fill up the
discrepancy, a new physics beyond the standard model is called
for. The apparent deviation is comparable to or even larger than the
contribution from the standard model electroweak sector computed as $
a_{\mu}(\mbox{SMEW}) =15.1(0.4) \times 10^{-10}$ \cite{MDMSMEW}. This
suggests that the energy scale of the new physics should be very close
to the electroweak scale and/or it should have some enhancement
mechanism to give a large contribution to $a_{\mu}$.

Among other things, supersymmetry (SUSY) is the most promising
candidate for such a new physics. To solve the naturalness problem in
the Higgs sector of the standard model, the superparticle masses
should lie below the TeV scale. The SUSY contribution to $a_{\mu}$ has
been investigated in the literatures (see
\cite{LNW,UN,Moroi,CGW,FengMoroiAMSB,GOS,Blazek,CGR} and references
therein).\footnote{See Ref.\ \cite{NatYam,Gra} for scenarios of large
extra dimensions.} Generically it is sizable for superparticles
weighing less than 1 TeV.  And as we will explain shortly, it is
enhanced for large $\tan \beta$ region, where $\tan \beta$ is the
ratio of the vacuum expectation values (VEVs) of the two Higgs bosons
in the SUSY standard model.

The purpose of this paper is to reexamine the contribution to
$a_{\mu}$ in the framework of the {\em unconstrained} minimal supersymmetric
standard model (MSSM) in the light of the recently reported
experimental data. The unconstrained MSSM has more than 100 SUSY
breaking parameters and one usually impose some relations among the
model parameters: otherwise it would be very difficult to trace all
dependence of the parameters for some specific processes. The SUSY
contribution to $a_{\mu}$, however, depends only on seven MSSM
parameters as we will list below, and thus we can leave them as free
parameters to analyze their dependence.  Another important point that
makes the model independent analysis possible is that, unlike flavor
changing neutral current (FCNC) processes, we do not rely on
particular mechanisms to suppress SUSY FCNC to compute the SUSY
contribution to $a_{\mu}$.  Conclusions we will draw are therefore
very general.

The apparent deviation from the standard model prediction implies a
non-vanishing SUSY contribution.  We will identify the parameter region
of the MSSM which is capable to account for the discrepancy
(\ref{eq:MDMdeviation}) at 2-$\sigma$.  Explicitly we require the SUSY
contribution to lie in the following range,
\begin{equation}
    11 \times 10^{-10} < a_{\mu}(\mbox{SUSY}) < 75 \times 10^{-10}.
    \label{eq:MDMSUSY}
\end{equation}
The SUSY contribution to $a_{\mu}$ decreases as the superparticles
become heavier.  Since a non-vanishing SUSY contribution is needed, we
will obtain an upper bound on the mass scale of the superparticles.
For large $\tan\beta$ region, the enhancement mechanism works and
(\ref{eq:MDMSUSY}) is easily satisfied even for relatively large
superparticle masses.  We will find that the Wino as heavy as 1 TeV
can be compatible with it.  Another point we pay a particular
attention is the case of low $\tan \beta$.  We will show that for
$\tan\beta \gtrsim 3$ the SUSY contribution can be large enough to
explain the deviation, without confronting the present bounds on the
superparticle masses obtained by negative searches of superparticles
at collider experiments.

At one-loop level, the SUSY contribution to $a_{\mu}$ stems from
chargino-sneutrino loops as well as from neutralino-smuon loops.
Formulae of the SUSY loop contributions are given, for example, in
Ref.~\cite{Moroi}.  In the (unconstrained) MSSM, the parameters
involved with this process are the soft SUSY breaking mass parameters
for the right-handed and left-handed smuons denoted by
$m_{\tilde{\mu}_R}$ and $m_{\tilde{\mu}_L}$, respectively, the
trilinear scalar coupling for muon $A_{\mu}$, the $U(1)_Y$ and
$SU(2)_L$ gaugino mass parameters $M_1$ and $M_2$, the higgsino mass
parameter $\mu$, and $\tan\beta$ which is the ratio of the VEVs of two
Higgs bosons.  Throughout this paper, we take the parameters to be
real, and do not include possible CP phases which are known to be
generically small due to stringent limits on electric dipole
moments. (See, however, Ref. \cite{IbrNat}.) 
For most of the analysis we impose the GUT relation for
the gaugino masses $M_1: M_2 \approx 1:2$.  We checked that our
following results are almost unchanged even if the GUT relation is
relaxed.  In the MSSM analysis, we do not consider any particular SUSY
breaking scenario, and thus we do not introduce specific relations
among the remaining six parameters mentioned above.  For comparison,
we will briefly discuss the case of the minimal supergravity scenario
(mSUGRA) later on. Furthermore we do not impose the lightest
superparticle (LSP) in the MSSM sector to be neutral. A charged LSP
would be ruled out if it were stable, but there are many ways out,
including R-parity violation and light gravitino LSP.

For generic SUSY mass parameters, it is known that the
chargino-sneutrino diagram gives a dominant contribution to
$a_{\mu}(\mbox{SUSY})$. Then the relevant parameters to calculate
$a_{\mu}(\mbox{SUSY})$ are $m_{\tilde \mu_L}$, $M_{2}$, $\mu$ and
$\tan\beta$. Thus it will be reasonable to fix the other mass
parameters for some specific values for the moment. Here we take
$A_{\mu}=0$ and assume the GUT relation to the gaugino
masses.\footnote{We checked that the value of $a_\mu(\mbox{SUSY})$ is
insensitive to the value of $A_{\mu}$ unless $A_{\mu}$ is extremely
large.} The sign of the SUSY contribution to $a_{\mu}$ is directly
correlated with the sign of $M_2 \mu$ in most of the parameter
regions. It is positive (negative) for $M_2 \mu >0$ ($M_2 \mu <0$).
Thus, the result of the E821 experiment given in Eq.\ 
(\ref{eq:MDMexp}) suggests $M_2 \mu >0$, and hence we consider the
positive sign case in the following.

Another important point is that the SUSY contribution to $a_{\mu}$ is
enhanced for large $\tan\beta$ because, in the dominant diagrams, muon
chirality is flipped by the muon Yukawa coupling
$y_{\mu}\propto1/\cos\beta\sim\tan\beta$, not by the muon mass itself
\cite{LNW,UN,Moroi}.  Thus, for the large $\tan\beta$ case, the SUSY
contribution can be large enough to explain the discrepancy even with
relatively heavy superparticles.  On the contrary, when $\tan\beta$ is
small, Wino and slepton masses should be light to make the SUSY
contribution to $a_{\mu}$ large enough.

\begin{figure}
\centerline{\epsfxsize=0.75\textwidth\epsfbox{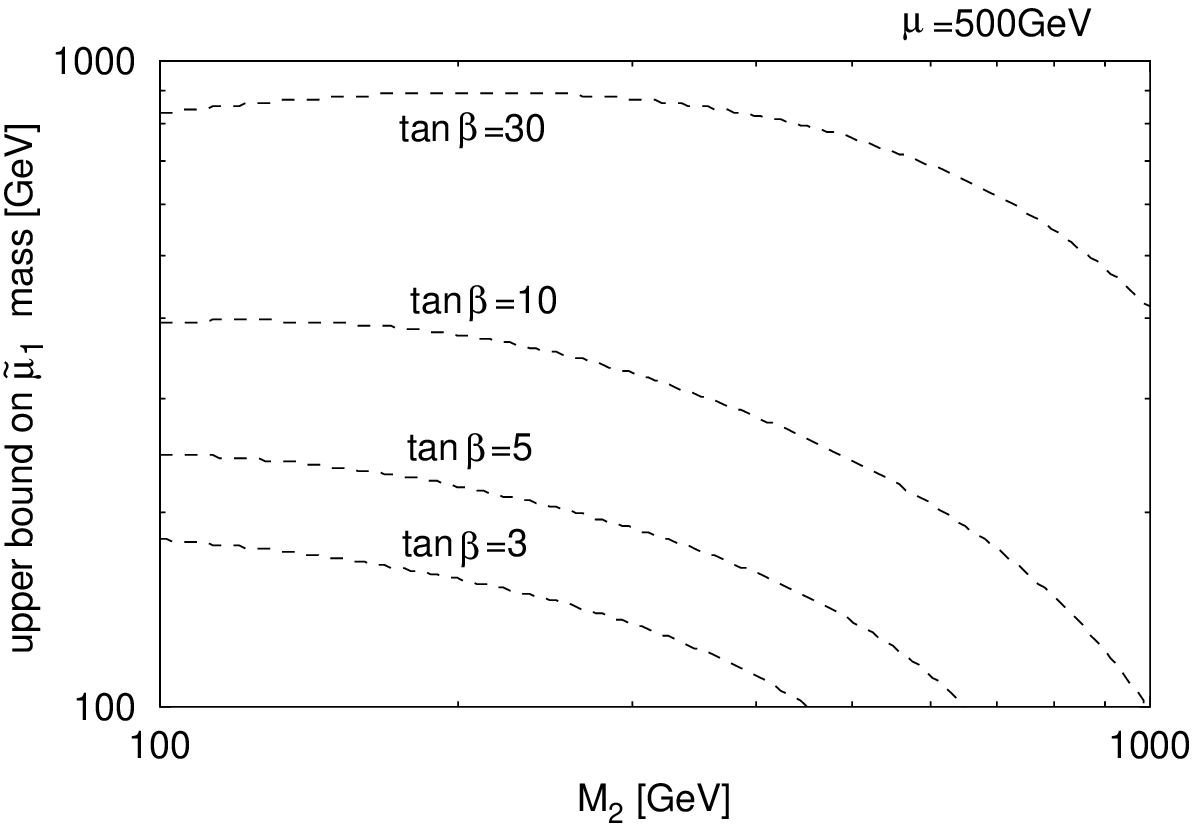}}
    \caption{The upper bounds on the lighter smuon mass
    $m_{\tilde{\mu}1}$. The horizontal line is the Wino mass parameter
    $M_2$, and we take $\mu=500\ {\rm GeV}$ and $A_\mu=0$.  Here,
    $\tan\beta$ is taken to be 3, 5, 10, and 30 from below.}
    \label{fig:MSSM_M2}
\end{figure}

In the framework of the unconstrained MSSM, we calculate an upper
bound on the lighter smuon mass $m_{\tilde{\mu}1}$ as a function of
the Wino mass parameter $M_2$.  The result is shown in Fig.\ 
\ref{fig:MSSM_M2}.  In deriving the upper bound, we vary $m_{\tilde
\mu _R}$ and $m_{\tilde \mu_L}$, and derive the largest possible value
of $m_{\tilde{\mu}1}$ which can realize the 2-$\sigma$ bound
(\ref{eq:MDMSUSY}).  In Fig.\ \ref{fig:MSSM_M2}, we take
$\tan\beta=3$, 5, 10, 30, and we fixed $\mu=500$ GeV as a typical
example.  We checked that the lighter smuon is mostly the left-handed
one which consists of the $SU(2)_L$ doublet with the sneutrino as it
is involved with the dominant chargino-sneutrino loop.  We find from
this figure that quite a large parameter region is in accord with the
consideration of the recent data of $a_{\mu}$. Thus we conclude that
the SUSY is naturally able to explain the apparent discrepancy
observed at the experiment.

Fig. \ref{fig:MSSM_M2} also shows that the constraint on the lighter
smuon mass $m_{\tilde{\mu}1}$ is much stronger than that on the Wino
mass $M_2$.  For instance, for $\tan\beta=10$, the smuon must be
lighter than about 400 GeV while the Wino as heavy as 1 TeV is allowed
at 2-$\sigma$.

\begin{figure}
\centerline{\epsfxsize=0.75\textwidth\epsfbox{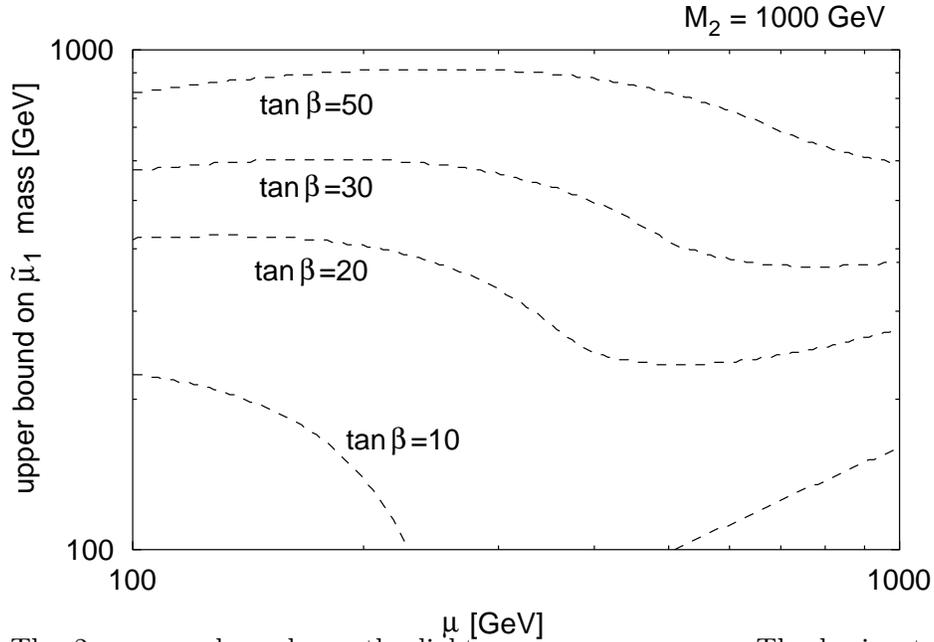}}
    \caption{The 2-$\sigma$ upper bounds on the lighter smuon mass
    $m_{\tilde{\mu}1}$. The horizontal line is the Higgsino mass
    parameter $\mu$, and we take $M_2=1\ {\rm TeV}$, $A_\mu=0$, and
    $\tan\beta=10$, 20, 30, 50.}
    \label{fig:MSSM_mu}
\end{figure}

Let us now closely look at the large $\tan\beta$ case. In this case,
the SUSY contribution is enhanced as $\tan\beta$ increases. Thus one
can expect that even heavy superparticles can be compatible with the
lower bound $a_{\mu}(\mbox{SUSY}) =11 \times 10^{-10}$. In Fig.\ 
\ref{fig:MSSM_mu}, we show a plot of the 2-$\sigma$ upper bounds on
$m_{\tilde{\mu}1}$ as a function of the $\mu$ parameter.  Here we take
several values of $\tan\beta$ and $M_2=1\ {\rm TeV}$.  Even with such
a large Wino mass, we find that the 2-$\sigma$ constraint can be
satisfied with $\tan\beta$ as small as 10 when the slepton mass is
lighter than about 200 GeV.  Applying the GUT relation of the gaugino
masses to the gluino mass as well, the Wino mass $M_2=1\ {\rm TeV}$
corresponds to the gluino mass of about 3 TeV. Namely the accuracy of
the present $a_{\mu}$ data still allows the possibility of such a
heavy gluino in the framework of the unconstrained MSSM. The Large
Hadron Collider (LHC) experiment will not be able to reach such a
heavy gluino, and thus SUSY searches at the LHC would require
unconventional approaches.  A future analysis of $a_{\mu}$ with more
statistics will further reduce the error in the measurement, which may
constrain the superparticle masses within the reach of the LHC.

Let us turn to the case of low $\tan\beta$. In this case the
enhancement mechanism operated at large $\tan\beta$ is not effective
and thus one may expect that the SUSY contribution is rather small.
The case of $\tan\beta=3$ in Fig. 1 shows, however, that the SUSY
contribution can explain the deviation at 2-$\sigma$ level when the
SUSY mass parameters are close to their experimental bounds.\footnote
{Existence of light superparticles may affect the fit to the
electroweak precision data, which was considered in Ref.\ 
\cite{ChoHagHay} in the parameter region where $a_{\mu}(\mbox{SUSY})$
is sizable.  According to Ref.\ \cite{ChoHagHay}, there is some region
of light higgsino dominant LSP case in which the inclusion of SUSY
particles gives a better fit to the electroweak precision data than the
standard model alone.}

Next we would like to discuss the mass of the lightest scalar Higgs
boson $m_h$. It is known that small $\tan\beta$ tends to give
relatively light $m_h$ because the tree-level contribution to $m_h^2$
is roughly given as $m_Z^2\cos^2 2\beta$. To survive the Higgs mass
bound obtained at LEP 200, which approaches 113.5 GeV as the
pseudo-scalar Higgs mass increases \cite{LEPHiggs}, large radiative
correction from top and stop loops is required \cite{HiggsRC}. Thus
the stop masses should be large enough.  Roughly speaking, the mass
bound is satisfied for small $\tan \beta \approx 3$ if the stop masses
exceed about 1 TeV.  The relation between the slepton masses which are
constrained by the $a_{\mu}$ analysis and the squark masses is highly
model dependent.

\begin{figure}
\centerline{\epsfxsize=0.75\textwidth\epsfbox{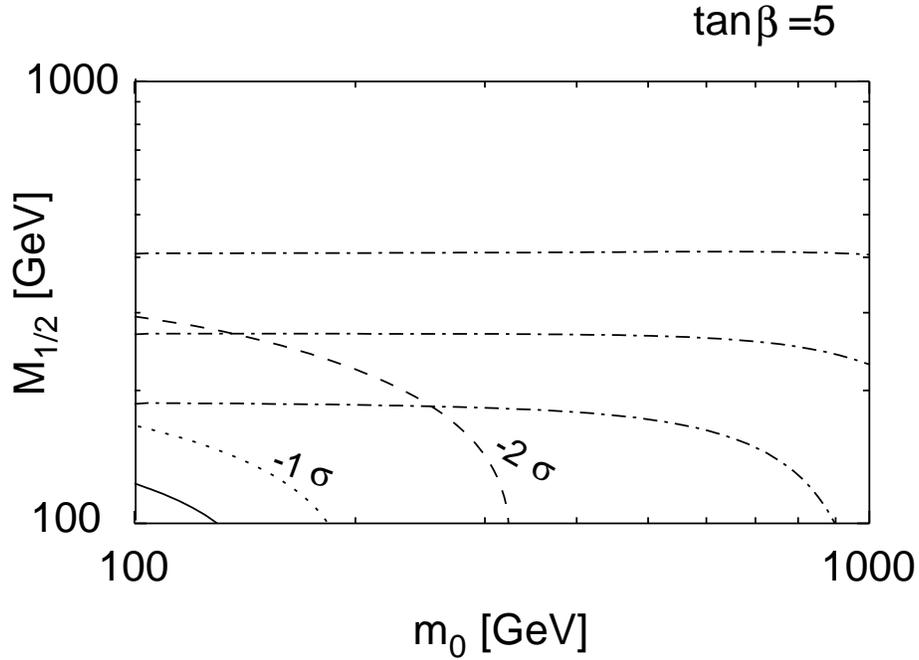}}
    \caption{Contours of the constant $a_\mu(\mbox{SUSY})$ on $m_0$
    vs.\ $M_2$ plane in the minimal supergravity model.  The solid
    (dotted, dashed) line is for the center ($\pm 1$-$\sigma$, $\pm
    2$-$\sigma$) value of $a_\mu(\mbox{SUSY})$.  We take $A_\mu=0$ and
    $\tan\beta=5$.  We also plot the constant $m_h$ contours in the
    dash-dotted lines ($m_h=110$, 115, and 120 GeV from below).}
    \label{fig:SUGRA_tb05}
\end{figure}

\begin{figure}
\centerline{\epsfxsize=0.75\textwidth\epsfbox{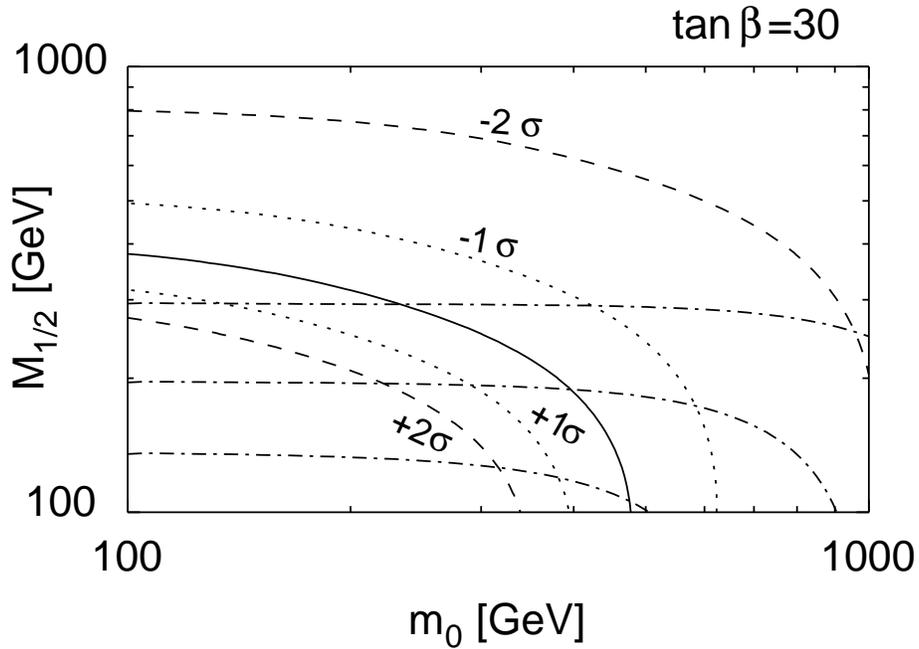}}
    \caption{Same as Fig.\ \ref{fig:SUGRA_tb05}, except $\tan\beta=30$.}
    \label{fig:SUGRA_tb30}
\end{figure}

To give more quantitative arguments on the light Higgs mass, we
consider the case of the mSUGRA.  In Figs.\ \ref{fig:SUGRA_tb05} and
\ref{fig:SUGRA_tb30}, we plotted contours of 0-$\sigma$, 1-$\sigma$,
and 2-$\sigma$ preferred values of $a_\mu(\mbox{SUSY})$ on $m_0$ vs.\ 
$M_2$ plane, where $m_{0}$ is the universal scalar mass at the GUT
scale and $M_{1/2}$ is the universal gaugino mass at the same
scale. (The Wino mass at the electroweak scale is related to $M_{1/2}$
as $M_2 \approx 0.83 M_{1/2}$.)  In this framework, we also calculated
the lightest Higgs mass $m_h$, and plotted the contours of constant
$m_h$ on the same figures.  In the figures the trilinear scalar
coupling is taken to be zero.  As one can see, for the case of
$\tan\beta=5$, the constraint $m_h\geq 113.5\ {\rm GeV}$ severely
constrains the parameter region which gives preferred value of
$a_\mu(\mbox{SUSY})$.  Taking account of the Higgs mass constraint,
however, the 2-$\sigma$ constraint (\ref{eq:MDMSUSY}) can be realized
with $\tan\beta\gtrsim 5$.  It is interesting to see that the region
with small universal scalar mass is favored.  Although the region with
$m_0< 100\ {\rm GeV}$ is not shown in the figures, we checked that
2-$\sigma$ constraint and the Higgs mass constraint can be
simultaneously satisfied in the limit of $m_0\rightarrow 0$ (i.e.,
with the no-scale type boundary condition).  For smaller $\tan\beta$,
$a_\mu(\mbox{SUSY})$ cannot be large enough to explain the discrepancy
at the 2-$\sigma$ level.  For large $\tan\beta$, on the contrary, the
Higgs mass can be easily large enough in the region with sufficient
$a_\mu(\mbox{SUSY})$.

Here, let us comment on the allowed parameter region in the mSUGRA
case.  We find that the allowed parameter region is rather tighten in
the mSUGRA scenario. To illustrate this point, let us consider the
case of $\tan\beta=5$. In the unconstrained MSSM, the 2-$\sigma$
bound constrains $M_2\lesssim 600$ GeV.  On the other hand, in the
mSUGRA case, larger value of $M_{1/2}$ results in larger slepton mass
through renormalization group effect.  As a result, the 2-$\sigma$
bound in the mSUGRA case is $M_{1/2}\lesssim 300$ GeV which
corresponds to $M_2\lesssim 250$ GeV.\footnote{We expect that a similar
tight bound on $M_2$ can be obtained in a wide class of models where the
supersymmetry breaking effect is mediated at high energy scale, and thus
the left-handed smuon mass suffers from the renormalization effect of 
$SU(2)_L$ gaugino.}  The difference will be crucial
when one considers the discovery potential of superparticles at hadron
colliders.

\begin{figure}
\centerline{\epsfxsize=0.75\textwidth\epsfbox{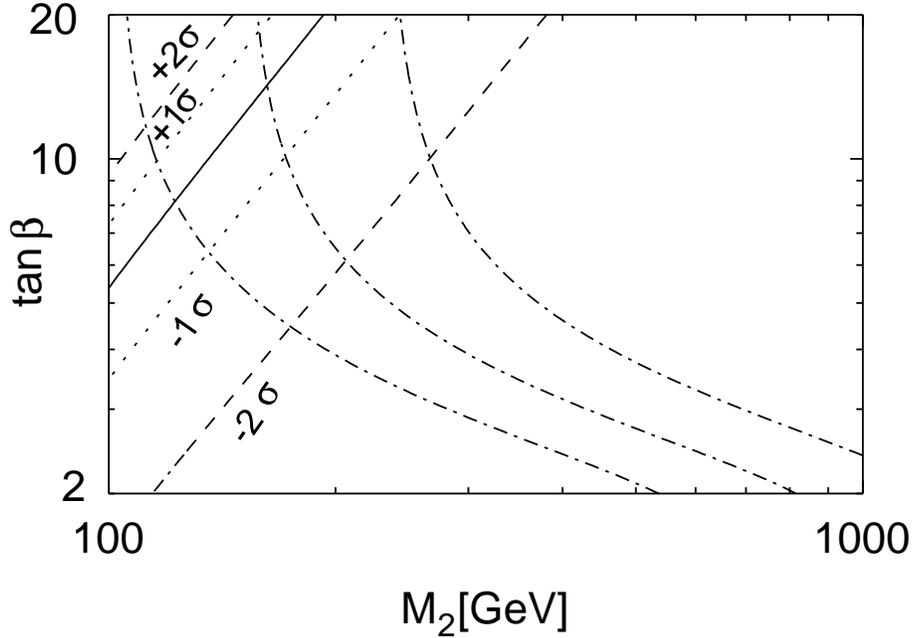}}
    \caption{Contours of the constant $a_\mu(\mbox{SUSY})$ on $M_2$
    vs.\ $\tan\beta$ plane in the gauge-mediated SUSY breaking
    scenario.  The solid (dotted, dashed) line is for the center ($\pm
    1$-$\sigma$, $\pm 2$-$\sigma$) value of $a_\mu(\mbox{SUSY})$.  We
    take $M_{\rm mess}=10^{6}\ {\rm GeV}$ and $N_{\rm mess}=1$. We
    also plot the constant $m_h$ contours in the dash-dotted
    lines. ($m_h=110$, 115, and 120 GeV from below).}
    \label{fig:GMSB}
\end{figure}

To make a comparison, we also consider a case of the gauge-mediated
SUSY breaking (GMSB) scenario whose free parameters are the overall
scale of the soft-breaking parameters $\Lambda_{\rm GM}$, the
messenger scale $M_{\rm mess}$, $\tan\beta$, and the number of the
${\bf 5}+{\bf \bar{5}}$ messenger multiplets $N_{\rm mess}$
\cite{GiuRat}.  The overall scale is related to the Wino mass as
\begin{eqnarray}
    M_2 = \frac{N_{\rm mess} g_2^2}{16\pi^2} \Lambda_{\rm GM},
\end{eqnarray}
with $g_2$ being the $SU(2)_L$ gauge coupling constant.  In our
analysis, we use $M_2$ instead of $\Lambda_{\rm GM}$ using this
relation.  In Fig.\ \ref{fig:GMSB}, we show the region which is
consistent with the 1-$\sigma$ and 2-$\sigma$ constraint on $M_2$ vs.\ 
$\tan\beta$ plane.  Here, we take $M_{\rm mess}=10^{6}\ {\rm GeV}$ and
$N_{\rm mess}=1$.  In addition, we also plot the constant $m_h$
contour.  We can see that $a_\mu(\mbox{SUSY})$ can be within the
2-$\sigma$ bound with $\tan\beta\gtrsim 5$ even with the Higgs mass
constraint $m_h\geq 113.5\ {\rm GeV}$.  We checked that this result is
insensitive to the choices of $M_{\rm mess}$ and $N_{\rm mess}$.

Finally, let us briefly comment on the branching ratio of
$b\rightarrow s\gamma$.  To this process, the SUSY contribution is
dominated by the stop-chargino loops. It is well-known that it can
interfere with the standard model and charged Higgs contribution
constructively or destructively, depending on the relative sign of the
$\mu$ parameter and the trilinear scalar coupling of stop $A_t$.
Furthermore, in most cases the sign of $A_t$ is essentially determined
by the sign of the gluino mass $M_3$ via large renormalization group
effect. It is found that for $M_3 \mu<0$ the SUSY contribution is
constructive, which is in fact disfavored because the branching ratio
$B(b \rightarrow s \gamma)$ tends to be predicted too large compared
to the experimental value.  Thus the $b \rightarrow s \gamma$
consideration favors the $M_3 \mu>0$ case. On the other hand, the
positive contribution to the $a_{\mu}$ is obtained when the sign of
$M_2 \mu$ is positive, provided that the chargino loop dominates over
the neutralino loops. Thus the case that the Wino and gluino have
masses with the same sign is favored by the combined consideration of
$a_{\mu}$ and $B(b \rightarrow s \gamma)$. This is the case in many
models of SUSY breaking. In particular, models with the GUT relation
of the gaugino masses fall into this category, including the mSUGRA
and the GMSB.  On the other hand, the case with the opposite sign of
the Wino and gluino masses is disfavored, which is in fact the case in
a simple class of anomaly mediated supersymmetry breaking models
\cite{FengMoroiAMSB}.

In this paper, we considered implications of the recently reported
$a_{\mu}$ measurement at the E821 experiment to supersymmetric
standard models. We made the analysis mainly based on the
unconstrained MSSM. Since the SUSY contribution should be
non-vanishing at 2-$\sigma$ level, the upper bounds on the
superparticle masses were obtained.  For large $\tan \beta$, the
superparticles can be quite heavy, which may be escaped from the LHC
gluino reach. We observed that the allowed region of the
superparticles masses is significantly larger than the case of
constrained models such as the mSUGRA model.  On the contrary for
smaller $\tan \beta$, the Wino mass as well as the lighter slepton
mass should be light. In this case, the bound on the Higgs boson mass
obtained at LEP200 gives stringent constraints. We illustrated this
point in the mSUGRA model as well as the GMSB model, yielding $\tan
\beta \gtrsim 5$ for both models.

Given the upper bounds on the masses of the sleptons and the
charginos/neutralinos, lepton flavor violation such as $\mu\rightarrow
e\gamma$ as well as $\mu$-$e$ conversion may be observed in near
future experiments \cite{LFV}.  Another implication is to proton
decay.  As we discussed, a model dependent lower bound on $\tan\beta$
is obtained by combining analyses of $a_{\mu}(\mbox{SUSY})$ and $m_h$.
This tightens proton decay constraints in SUSY GUT models, though
details are quite model dependent.  Result of detailed study along
this line will be presented elsewhere.

To conclude, we should emphasize the importance of the further
reduction of error of the $a_{\mu}$ measurement which is expected to
be done in near future, as well as the further study of the
uncertainty coming from the hadronic vacuum polarization.  We hope
that they may sharpen the confrontation with the standard model more
clearly in near future and confirm the necessity of the physics
beyond the standard model.

While preparing the manuscript, we were aware of the papers:
A. Czarnecki and W.J. Marciano, hep-ph/0102122,
L. Everett, G.L. Kane, S. Rigolin and L.-T. Wang, hep-ph/0102145,
J.L. Feng and K.T. Matchev, hep-ph/0102146,
E.A. Baltz and P. Gondolo, hep-ph/0102147,
U. Chattopadhyay and P. Nath, hep-ph/0102157,
which have some overlap with our analyses.

{\sl Acknowledgment:} This work was supported in part by the
Grant-in-aid from the Ministry of Education, Culture, Sports, Science
and Technology, Japan, priority area (\#707) ``Supersymmetry and
unified theory of elementary particles,'' and in part by the
Grants-in-aid No.11640246 and No.12047201.

\end{document}